\documentstyle[12pt]{article}
\begin{document} 
\begin{center} 
%\begin{flushleft} 
{\bf HST OBSERVATIONS OF THE BLUE COMPACT DWARF
 SBS 0335--052: A PROBABLE YOUNG GALAXY
{\footnote[1]{Based on observations 
obtained with the NASA/ESA {\it{Hubble Space Telescope}} through the Space 
Telescope Science Institute, which is operated by AURA,Inc. under NASA
 contract 
NAS5-26555}}}

\vspace{0.5cm}
{\bf Trinh X. Thuan}%{\footnote[1]{Visiting Astronomer, Kitt Peak National 

{\small\rm Astronomy Department, University of Virginia, Charlottesville
VA 22903, USA; txt@virginia.edu}

{\bf Yuri I. Izotov} 
 
{\small\rm Main Astronomical Observatory, Ukrainian Academy of Sciences, 
 
Goloseevo 252650, Kiev-22, Ukraine; izotov@mao.gluk.apc.org}

and 
 
{\bf Valentin A. Lipovetsky} %{\footnotemark[1]}}

{\small\rm Special Astrophysical Observatory, Russian Academy of Sciences, 
 
Nizhny Arkhyz, Karachai-Circessia 357147, Russia; val@sao.stavropol.su} 
 
\vspace{1.0cm} 
%\vspace{0.5cm} 
 
{\sl{Received}\underline{\ \ \ \ \ \ \ \ \ \ \ \ \ \ \ \ \ \ \ \ \ \ \ \ \ } 
\sl{Accepted}\underline{\ \ \ \ \ \ \ \ \ \ \ \ \ \ \ \ \ \ \ \ \ \ \ \ \ }  
} 
 
\vspace{1.cm} 
Submitted to the {\sl Astrophysical Journal} 
%\end{flushleft} 
\end{center} 
\newpage 
\begin{center} 
ABSTRACT 
\end{center}

  We present HST WFPC2 V and I images and GHRS UV spectrophotometry of 
the spectral regions around Ly$\alpha$ and OI $\lambda$1302 of the 
extremely metal-deficient ( Z $\sim$ Z$_\odot$/41 ) blue compact dwarf 
(BCD) galaxy SBS 0335--052. 
 All the star formation in the BCD occurs in six super-star clusters
(SSC) with ages $\leq$ 3-4 Myr. As there is no evident sign of 
 tidal interaction, star formation 
in the BCD is probably triggered by stochastic cloud collisions in the 
HI envelope. Dust is clearly present and mixed spatially with the SSCs.
There is a supershell of radius $\sim$ 380 pc, delineating a large 
supernova cavity. The instantaneous star formation rate is $\sim$ 0.4 
M$_\odot$ yr$^{-1}$ .

 Strong narrow Ly$\alpha$ emission is not observed. Rather there is  
low intensity broad (FWZI = 20 \AA) Ly$\alpha$ emission superposed on even 
broader Ly$\alpha$ absorption by the HI envelope. This broad low-intensity
emission 
is caused by resonant scattering of Ly$\alpha$ photons. The absence of 
strong Ly$\alpha$ emission may be due partly to dust absorption,
 but mainly to 
multiple scattering which removes Ly$\alpha$ photons from the small HST 
aperture. As the HI cloud is seen nearly edge-on, geometrical effects 
may play also a role as photons escape more easily in a direction 
 perpendicular to the plane than along it.

The BCD appears to be a young galaxy, undergoing 
its very first burst of star formation. This conclusion is based on the 
following evidence: 1) the underlying extended low-surface-brightness
component  is very 
irregular and filamentary, suggesting that a  significant part of the  
emission comes from ionized gas;
%not from hot stars;
2) it has very blue colors (--0.34 $\leq$ (V--I)$_0$ $\leq$ 0.16), consistent 
with gaseous emission colors; 
 3) the OI $\lambda$1302 line is not detected 
in absorption in the GHRS spectrum, setting an upper limit for N(O)/N(H) 
in the HI envelope of the BCD of more than 3000 times smaller than the 
value in Orion.
   
{\it{Subject headings:}} galaxies -- young: interstellar matter -- nebulae: 
HII regions 

\newpage 
\begin{center} 
1. INTRODUCTION 
\end{center}

      The formation of galaxies is one of the most fundamental problems in
astrophysics, and much effort has gone into the search for primeval galaxies
(PG). A possible definition of a primeval galaxy is a young system undergoing
its first major burst of star formation. It is now widely believed that the
vast majority of galaxies underwent such a phase at redshifts $\sim$2 or
greater. In most galaxy formation scenarios, young galaxies are predicted to
show strong Ly$\alpha$ emission, associated with the cooling of the primordial
gas and the subsequent formation of a large number of massive ionizing stars
(Partridge \& Peebles 1967; Meier 1976; Baron \& White 1987; Charlot \& Fall
1993). Yet, despite intensive searches, the predicted widespread population of
Ly$\alpha$ primeval galaxies has remained elusive (Pritchett 1994). The reasons
for this non-detection are not yet clear.

      Several objects have been put forward as possible PG candidates, ranging
from high-redshift radio galaxies to Ly$\alpha$ emitters found around quasars
and damped Ly$\alpha$ systems, mainly on the basis of very high luminosity and
star formation activity (Fontana et al. 1996; Pettini, Lipman \& Hunstead 1995;
Yee et al. 1996). However, most of these candidate PGs already contain a
substantial amount of heavy elements, implying previous star formation and
metal-enrichment and hence not satisfying the above definition of PG. For
example, the spectrum of the PG candidate discussed by Yee et al. (1996) shows
a strong P-Cygni profile in CIV $\lambda$1550, indicating the presence of heavy
elements since low-metallicity stars do not have strong winds and show no or
weak P-Cygni profiles. Moreover, even if PG candidates at high-redshift are
discovered, it is difficult to study them in detail because of their extreme
faintness and very compact angular size. Here we take a different approach to
the PG problem. Instead of searching for very high-redshift galaxies in the
process of forming, we look for nearby galaxies undergoing their first burst of
star formation, and hence satisfying the above definition of a PG. The best
candidates for such a search are blue compact dwarf galaxies (BCD).

      BCDs are extragalactic objects with M$_B$$\geq$--18 where intense star
formation is presently occuring, as evidenced by their blue UBV colors, and
their optical spectra which show strong narrow emission lines superposed on a
stellar continuum which is rising toward the blue, similar to spectra of HII
regions. Star formation in BCDs cannot be continuous but must proceed by bursts
because of several observational constraints: 1) Gas is transformed into stars
at the rate of approximately 1 M$_\odot$yr$^{-1}$, so that the current burst
cannot last more than about 10$^8$ yr before depleting the gas supply; 
2) Optical-infrared colors of BCDs give ages of about 10$^7$ yr; and 3) 
Population synthesis of UV spectra of BCDs give invariably jumps in the
stellar luminosity function, indicative of starbursts (see Thuan 1991 for a
review).

      Ever since their discovery, the question has arisen whether BCDs are 
truly young systems where star formation is occuring for the first time, or old
galaxies with an old underlying stellar population on which the current 
starburst is superposed (Searle, Sargent \& Bagnuolo 1973). Thuan (1983)
 carried
out a near-infrared JHK survey of BCDs and concluded that all the objects in
 his
sample possessed an old underlying stellar population of K and M giants. That
result was not unambiguous as the JHK observations were centered on the
star-forming regions and the near-infrared emission could be contaminated by
light from young supergiant stars. The advent of CCD detectors allowed to look
for the low-surface-brightness underlying component directly. Loose \& Thuan
(1985) undertook a CCD imaging survey of a large BCD sample and found that
nearly all galaxies ($\geq$ 95\%) in their sample show an underlying extended 
low-surface-brightness component, on
which are superposed the high-surface-brightness star-forming regions.
Subsequent CCD surveys of BCDs have confirmed this initial result (Kunth, 
Maurogordato \& Vigroux 1988, Papaderos et al. 1996). Thus, 
most BCDs are not young galaxies. In Kunth et al. (1988)'
sample, the BCD I Zw 18 does not show an underlying component. This
BCD with a metallicity of only Z$_\odot$/50 is the most metal-deficient galaxy
known (Searle \& Sargent 1972). It has been discussed as a possible young
galaxy, currently undergoing its first burst of star formation (Kunth \&
Sargent 1986). Hunter \& Thronson (1995) obtained HST images of I Zw 18 and
concluded that the colors of the underlying diffuse component are consistent
with those from a sea of unresolved B or early A stars, with no evidence for
stars older than $\sim$ 10$^7$ yr.

      For more than 20 years, I Zw 18 stood in a class by itself. The BCD
metallicity distribution ranges from $\sim$Z$_\odot$/3 to $\sim$Z$_\odot$/50,
peaking at $\sim$Z$_\odot$/10, and dropping off sharply for Z$\leq$Z$_\odot$/10
(Kunth \& Sargent 1986). Intensive searches have been carried out to look for
low-metallicity BCDs but they have met until recently with limited success. For
example, the majority of the BCDs in the Salzer (1989) and Terlevich et al.
(1991) surveys have metallicities larger than Z$_\odot$/10. Several years ago,
a new BCD sample has been assembled by Izotov et al. (1992, 1993) from 
objective prism survey plates obtained with
the 1m Schmidt telescope at the Byurakan Observatory of the Armenian Academy
of Sciences during the Second Byurakan Survey (SBS, Markarian, Lipovetsky \&
Stepanian 1983). The most interesting feature of the SBS is its metallicity
distribution (Izotov et al. 1992): it contains
significantly more low-metallicity BCDs than previous surveys. It has uncovered
about a dozen BCDs with Z$\leq$Z$_\odot$/15, more than doubling the number of 
such known low-metallicity BCDs and filling in the metallicity gap between
I Zw 18 and previously known BCDs. This low-metallicity sample has been used to
determine the primordial helium abundance (Izotov, Thuan \& Lipovetsky 1994,
1996), heavy element abundances in metal-deficient environments (Thuan,
Izotov \& Lipovetsky 1995) and study the BCD large-scale spatial distribution
 (Pustilnik et al. 1995).

 Here we focus our attention on the most 
metal-deficient BCD in the SBS sample, SBS 0335--052. Using the 
 Hubble Space Telescope (HST), we have obtained optical imaging
and UV spectroscopy for the galaxy.  We discuss the 
HST observations and their 
data reduction in \S2. We argue for the youth of 
SBS 0335--052 in \S3. In \S4 we show how other observational data also support 
the youth hypothesis, and  discuss the Ly$\alpha$ problem in the BCD and its 
implications on Ly$\alpha$ searches of high-redshift primeval galaxies. 
 We also compare 
the properties of SBS 0335--052 with those of other nearby young 
 galaxy candidates. 
 We summarize our conclusions in \S5.

\begin{center} 
2. OBSERVATIONS AND DATA REDUCTION 
\end{center}

      The BCD SBS 0335--052 was discovered by Izotov et al. (1990) to have an
extremely low metallicity. With Z$\sim$Z$_\odot$/41 (Melnick, Heydari-Malayeri
\& Leisy 1992), it is the second most-metal deficient BCD known, only behind
I Zw 18. Its integral characteristics are summarized in Table 1. At the
adopted distance{\footnote[2]{We shall adopt throughout a Hubble constant of
75 km s$^{-1}$Mpc$^{-1}$.}} of 54.3 Mpc, 1$''$ corresponds to a linear size of
263 pc. To understand better the nature of this extraordinary object, we have
obtained images and high resolution UV spectra of SBS 0335--052 
 with the refurbished HST , in the course of a larger study of 
 extremely metal-deficient BCDs . The superior angular resolution of HST is
 ideal for studying this 
extremely compact (d$_{25}$$\sim$12$''$) object.

\begin{center}
2.1. Imaging
\end{center}

      We obtained images of SBS 0335--052 on 1995 January 27 during cycle 4,
after the refurbishment mission, with the HST Wide Field and 
 Planetary Camera 2 
(WFPC2) in filters F569W and F791W, which we will refer to as V and I 
throughout the paper. Two exposures of equal duration were obtained in each 
filter to permit identification and removal of cosmic rays. The total exposure
time was 1800 s in V and 4400 s in I. The scale of the WFPC2 is 0.102$''$ per
pixel. 

      The data reduction followed the procedures described in Thuan, Izotov \&
Lipovetsky (1996a). Preliminary processing of the raw images including 
corrections for flat-fielding was done at the Space Telescope Science 
 Institute (STScI)
through the standard pipeline. Subsequent reductions were carried out at the
University of Virginia using IRAF
{\footnote[3]{IRAF: the Image Reduction and Analysis Facility is distributed by
the National  Optical Astronomical Observatories which are operated by the
Association of Universities for Research in Astronomy, Inc. (AURA), under
cooperative agreement with the National Science Foundation (NSF).}}
and STSDAS
{\footnote[4]{STSDAS: the Space Telescope Science Data Analysis System.}}.
Point sources in the processed images have a FWHM of $\sim$ 2 pixels or 
 0.2$''$.
Cosmic rays were removed and the images in each filter were combined. We found
that all exposures in a given filter coregistered to better than $\sim$ 0.2
pixels. The transformation of instrumental magnitudes to the Johnson-Cousins
UBVRI photometric system as defined by Landolt (1992) was performed according
to the prescriptions of Holtzman et al. (1995). The resulting measured 
brightness of the sky background is 22.6 mag arcsec$^{-2}$ in V and 21.8
mag arcsec$^{-2}$ in I.

\begin{center}
2.2. UV spectroscopy
\end{center}

Spectroscopic observations were obtained on 1995 January 3-4  using
the Goddard High Resolution Spectrograph (GHRS) on board HST, in two 
separate exposures, one with
 the G160M grating covering the 1210--1250\AA\ region around the Ly$\alpha$ 
line of hydrogen and the other with the same grating covering  
 the 1300--1350\AA\ region around the resonance line of oxygen at 1302 \AA.
 The exposure time was 7180 seconds for each region.
 The 2$''$$\times$2$''$ rectangular Large Science Aperture 
 was used, centered on the brightest knot (corresponding to cluster 1 in Table 
2 to be discussed later). The 
 flux and wavelength calibrations were done at STScI through the standard 
 pipeline. The spectral resolution is 0.12 \AA.

\begin{center} 
3. A PROBABLE YOUNG GALAXY
\end{center}

\begin{center}
3.1. {\sl Young super-star clusters} 
\end{center}

      We display in Figure 1a the V image with the contrast level adjusted so
as to show the details of the high-surface-brightness star-forming regions. The
latter are seen to contain 6 bright stellar clusters all within a region of 
$\sim$ 2 arcsec diameter, or $\sim$ 520 pc in linear size. It is clear from the
contour map (Figure 1b) that the clusters (labeled 1 to 6) are unresolved, with
radii $\leq$25 pc.
Aperture photometry of the point-like objects was performed using the package 
APPHOT in IRAF. The magnitudes were obtained using a circular aperture of
radius 2 pixels, which is about the FWHM of the stellar profile.
 To determine the background, we use an annulus centered on the cluster
covering 3--5 pixels. The precision of the photometry is good despite the high
background because of the relative brightness of the clusters
 (V$\leq$22 mag).

 Table 2
shows the V$_0$ and I$_0$ magnitudes, and the (V -- I)$_0$ colors for all six
clusters. All magnitudes and colors have been corrected for Galactic extinction
which is small in the direction of SBS 0335--052 (A$_V$=0.10 mag, A$_I$=0.06 
 mag,
Burstein \& Heiles 1982). The magnitudes of the stellar clusters vary between
V$_0$ = 19 mag and V$_0$ = 22 mag, which correspond to absolute magnitudes
 M$_{V_0}$
between --11.7 mag and --14.7 mag. The (V -- I)$_0$ colors are very blue for
all clusters, ranging between --0.61 and 0.31. In fact, the (V -- I)$_0$ color
for the bluest and brightest cluster (No.1 in Table 2) is bluer than that of 
 the
hottest main sequence O star ( --0.33 for a OV star with solar metallicity,
Bessell 1990). Models of young (age $<$3--4 Myr) stellar clusters 
 (Leitherer \& 
Heckman 1995) predict 
(V -- I)$_0$ in the range 0.0 -- 0.1 for a Salpeter or a Miller-Scalo-type 
initial mass functions (IMF), with an upper mass limit of 100 M$_\odot$ and a
metallicity of one tenth that of the Sun, in the two limiting cases of an 
instantaneous burst and a constant star-formation rate over a time interval of
3$\times$10$^6$yr. Thus cluster 1 is $\sim$ 0.7 mag too blue in (V--I)$_0$ as 
compared to the model predictions.
Several effects can be thought of which make (V -- I)$_0$ bluer. The 
most important
effect is the contamination by gaseous emission, especially in the 
 V band where several 
strong emission lines are present. To estimate this 
 contamination , we have used the spectroscopic observations of 
 Izotov et al. (1996) who
have measured for cluster 1 equivalent widths of 178 \AA, 199 \AA, 616 \AA\ 
 and 1138 \AA\ 
respectively for the H$\beta$, [OIII]$\lambda$4959, 5007 and H$\alpha$ lines.
By convolving the line fluxes with the transmission curve for the V filter
(HST WFPC2 Handbook 1995), we estimate the contribution of gaseous emission to
the V band to be $\sim$ 0.5 mag. The remaining small $\sim$ 0.2 mag difference
can be plausibly attributed to contamination by gaseous continuous free-free
and free-bound emission and by a metallicity difference between models and
observations. Stellar clusters in SBS 0335--052 have Z$\leq$Z$_\odot$/41
(stars have a lower metallicity than the gas) while 
 Leitherer \& Heckman (1995)' models were computed for Z = Z$_\odot$/10. 
 A lower metallicity would result in a lower line blanketing, and hence 
to bluer colors. Thus we believe that the (V--I)$_0$ color
 of cluster 1 is consistent with the model calculations of
Leitherer \& Heckman (1995): the cluster was created in a starburst with
 a normal
IMF ($x$ = 1.35 where $x$ is defined by dN/d(logM) $\propto$ M$^{-x}$, M$_u$ =
100M$_\odot$, M$_l$=1M$_\odot$) less than $\sim$ 3--4 Myr ago.

 The clusters listed in Table 2 and 
 labeled in Figure 1b all lie roughly in a SE--NW direction, with cluster 1 at
the SE edge. They all have redder (V -- I)$_0$ colors than cluster 1.
 Figure 2 shows that there is a systematic increase in reddening
(although the decrease in brightness is not monotonic) of the clusters
 away from the brightest
cluster (No. 1). The color gradient $\Delta$(V -- I)$_0$/$\Delta$r is nearly
constant and equal to 
 $\sim$ 0.47
mag arcsec$^{-1}$ or $\sim$ 0.18 mag/100 pc. Thus cluster 6 which is
furthest away from cluster 1, at $\sim$2 arcsec or 526 pc, is $\sim$0.9 mag
redder. We argue that the reddening is not caused by evolutionary effects, but
rather by internal extinction by dust. The presence of dust patches is clearly
seen in Figure 3 where we show the (V -- I) color map of SBS 0335--052. Black
pixels denote blue colors while white pixels indicate red colors. Comparison
with Figure 1a shows that cluster 1 lies outside the dust patches, but that the
other clusters are in the same regions as the dust and are subjected to 
 internal
extinction. The dust patches are seen to systematically increase in importance 
 away from cluster 1, confirming visually the effect found in Figure 2. 
 There is quantitative evidence that internal extinction does increase 
 with
increasing distance from cluster 1. Melnick et al. (1992) have obtained with
the ESO New Technology Telescope (NTT) spectra of two knots in SBS 0335--052,
the first knot containing clusters 1 and 2 and the second knot clusters 4 and 
 5.
Although their seeing was excellent for ground-based observations (0.58$''$ --
0.85$''$ FWHM), thanks to the adaptive optics of the NTT, 
 Melnick et al. (1992) did not
have the necessary angular resolution to resolve those two knots into
individual clusters as done here with HST. Those authors found that the 
extinction coefficient C(H$\beta$) is only 0.07 dex for the first knot while it
is larger for the second one, C(H$\beta$) = 0.56 dex. This corresponds to a
reddening difference of E(B -- V) $\sim$ 0.3 mag, which is just about the 
amount needed to account for the color difference $\Delta$(V -- I)$_0$ = 0.45 
mag
between clusters 1 and 4 (Table 2). Thus there is good evidence
that most of the color gradient depicted in Figure 2 is not due to evolutionary
effects but to increasing extinction away from cluster 1, from A$_V$ = 0 for
cluster 1 to A$_V$ $\sim$ 1.8 mag for cluster 6.

      We now discuss the nature of the blue clusters in SBS 0335--052. After
correction for internal extinction, gaseous emission and metallicity effects so
that each cluster has the same intrinsic (V -- I)$_0$ $\sim$ 0.1 as cluster 1,
then the cluster absolute magnitudes are in the range --14.1 $\leq$
 M$_V$ $\leq$ 
--11.9. This is precisely the range of absolute magnitudes found by O'Connell,
Gallagher \& Hunter (1994) for objects called ``super-star clusters'' (SSC).
 SSCs are defined by their combination of 
 very
small sizes and high luminosities. They have diameters less than 50 pc, 
consistent with the radius upper limit of 25 pc set for our stellar clusters.
O'Connell et al. (1994) could resolve some of their objects with the HST
Planetary Camera and found that their half-light radii R$_{0.5}$ $\leq$ 3.5 pc,
comparable to radii of normal star clusters and compact HII regions. In spite
of their compactness, the luminosities of the SSCs are 
comparable to those of the physically larger superassociations (M$_V$ $\leq$
--11). This implies that SSCs have surface brightnesses $\geq$
100 times those of clusters and associations in normal giant HII regions. They
represent particularly intense periods of star formation where gas is 
 transformed into stars at the rate of $\sim$ 1M$_\odot$ 
 yr$^{-1}$
(Thuan 1991). From the H$\beta$ luminosity in knot 1 (clusters 1 and 2) of 
SBS 0335--052 measured by Melnick, Heydari-Malayeri \& Leisy (1992), 
we found that $\sim$ 10$^3$ --
10$^4$ massive O stars are present in that knot. Thus it appears that some 
 BCDs make 
stars by forming SSCs. This SSC formation appears to be related 
to the starburst mode of star formation, i.e., to a sharp increase in the star
formation rate of a galaxy during a relatively short ($\leq$ 10$^7$ yr) period.

 The advent of HST and its superior angular resolution has permitted the
discovery of SSCs with an ever greater frequency, all of them in galaxies
undergoing starbursts (see the review by Ho 1996). Examples include 
 the amorphous galaxy NGC 1140 (Hunter
et al. 1994), the BCD He2--10 (Conti \& Vacca 1994), the peculiar galaxy 
NGC 1275 (Holtzman et al. 1992), the merging systems NGC 7252 (Whitmore et al.
1993) and NGC 4038/4039 (Whitmore \& Schweizer 1995). Meurer et al. (1995) have
studied nine starburst galaxies ranging from BCDs to ultraluminous merging
far-infrared galaxies and also conclude that cluster formation is an important
mode of star formation in starburst galaxies. Even in galaxies where star
formation activity is not dominant, Barth et al. (1995) found SSCs within the
circumnuclear starburst rings around the Seyfert galaxies NGC 1097 and 
NGC 6951.

 It has been suggested (e.g. Meurer et al. 1995), mainly on the basis
 of sizes and luminosities, that these blue SSCs will fade out
to become globular clusters, once the starburst is over. We discuss this 
issue using our own HST data on stellar clusters in several BCDs in various 
evolutionary stages. Besides SBS 0335--052 which we shall argue, appears to be 
a young BCD undergoing star formation for the first time, we have also used
HST to study the BCD Markarian 996 which is at a much later stage of evolution,
having undergone several bursts of star formation before the present one
(Thuan, Izotov \& Lipovetsky 1996a). Projected against and around Mrk 996, are
many pointlike sources which can be interpreted \underline{not} as young but
old ($\geq$ 10$^{10}$ yr) globular clusters, as evidenced by their red color
((V--I)$_0$ $\sim$ 0.9), absolute magnitude range 
 (--10.4 $\leq$ M$_V$ $\leq$ --6.8)
and unresolved diameters ($\leq$ 20 pc). In the --9.0 $\leq$ M$_V$ $\leq$--7.5 
range, the cluster luminosity function in Mrk 996 is remarkably similar
to that of the Milky Way, with slightly more clusters in Mrk 996 relative
 to the
Galaxy in the magnitude range --10.5 $\leq$ M$_V$ $\leq$--9.0. The presence of 
red but not blue clusters around Mrk 996  implies that SSC formation 
does not always accompany a starburst in a BCD. If SSCs are indeed 
progenitors of globular clusters, SSC formation has occurred not with 
the present burst but with past ones in Mrk 996. Furthermore, the spatial 
distributions of clusters in the 2 BCDs are very different. In the case of 
SBS 0335--052, the SSCs are within the star-forming regions, while in the case 
of Mrk 996 the red globular clusters surround the galaxy, just as in the 
Milky Way.

Since SSC formation is associated with merging events, we have investigated 
the environment of SBS 0335--052. A VLA HI map of the BCD (Thuan et al. 1996b)
 shows 
 it to be embedded in a very large flattened HI cloud, 4.3$'$ EW by 1.5$'$ NS 
 in angular size or 64 kpc by 24 kpc in linear size. There are 
two peaks of HI column density, one of which is associated with 
SBS 0335--052 and the other with a fainter dwarf irregular galaxy (Pustilnik 
 et al. 1996).
% The HI map does not 
% show any evident sign of tidal interaction either. 
There is  a nearby companion, 
 the face-on Scd galaxy NGC 1376 ( m$_B$ = 12.8 mag, 2.0$'$$\times$1.7$'$ ) at 
 $\sim$9.5$'$ W [ $\alpha$(1950) = 03$^h$34$^m$37.2$^s$, 
$\delta$(1950)= --05$^{\circ}$12$'$ 29 $''$] and with a heliocentric 
velocity of 4159 km s$^{-1}$, 
 only 116 km s$^{-1}$ larger than the velocity of SBS 0335--052.
The VLA map shows no evident sign of tidal interaction between the HI
 cloud and 
NGC 1376. There are 
 three other galaxies with angular size $\sim$ 1$'$ within an angular 
 distance of 6.5$'$ from the BCD, but they are all background galaxies : 
 MPW1G at 4.3$'$ NE with v= 10779 km s$^{-1}$, A447 at 5.5$'$ SE with 
 v= 33660 km s$^{-1}$ and IRAS 03348--058 at 6.2$'$ NW with v= 5556 km 
 s$^{-1}$.  If NGC 1376 moves at 
 $\sim$ 100 km s$^{-1}$ relative to the BCD, it will take about 10$^9$ yr 
 to go the 150 kpc distance separating the two galaxies, much larger than 
 the $\leq$ 3--4 Myr age of the SSCs. Thus the formation of SSCs in 
SBS 0335--052  is probably not triggered by tidal interactions, but by 
 stochastic collisions of small HI gas clouds within the large HI envelope.

\begin{center}
3.2. {\sl A very blue underlying component}
\end{center}

Figure 1a shows that there is a more extended lower surface-brightness diffuse 
 component underlying the SSCs. This is seen better in figure 4 where the 
 contrast has especially been adjusted to display low-surface-brightness 
features. The underlying component 
 extends in the same SE--NW direction as the SSCs and is roughly 
 elliptical in shape, with the 
 SSCs located at the SE end. The nature of this underlying component is 
 crucial for determining the age of the galaxy, whether it is undergoing 
 its very first burst of star formation, in which case it satisfies the 
 definition of a primeval galaxy, or whether the present burst is just one 
 after many, in which case the underlying component contains an older 
 stellar population.

 We have derived surface brightness and color profiles for the underlying 
 component. Surface photometry was done by fitting ellipses to the isophotal 
 contours using the task ELLIPSE in STSDAS\footnotemark[4]. 
We show respectively in Figures 
 5 a,b and c the V and I surface brightness and (V--I) color distributions 
 as a function of log r. Here r is the equivalent radius of the best fitting 
 elliptical isophote defined as r = (ab)$^{1/2}$, where a and b are 
 respectively the semi-major and semi-minor axes of the ellipse. In Figures 
 5a and 5b, it can be seen that for r $\geq$ 1$''$, the surface brightness 
 distribution is well fitted by a power law (a straight line in this 
coordinate system) of slope 
 n $\sim$ 2.3 ( where n is defined as I $\propto$ r$^{-n}$) for 
 both V and I bands.
 The large statistical error bars in the radius range --0.45 $\leq$ log r 
 $\leq$ --0.05 or 0.4$''$ $\leq$ r $\leq$ 0.9$''$, is due to light 
 contamination from the shell structure seen clearly in Figure 4. This 
 structure is probably the result of supernova explosions in the SSCs 
 which have carved out a nearly spherical cavity in the ambient interstellar 
medium (ISM) of 
 the BCD.  As for the non-monotonically increasing error bars at radii 
 log r $\sim$ 0.3 or r $\sim$ 2$''$ and log r $\sim$ 0.7 or r $\sim$ 5 $''$, 
 they are probably due to light contamination from the SSCs at the SE side 
 of the galaxy. Most remarkable is the (V--I) color profile (Figure 5c) which 
 is nearly flat and shows extraordinary blue colors for the underlying 
 low-surface-brightness component. The (V--I) colors vary between --0.3 and 
 0.2, corresponding to foreground extinction corrected colors (V--I)$_0$ 
 between --0.34 and 0.16.

If the underlying emission is of stellar origin, this range of colors would 
 correspond to spectral types between OV and A5V stars, using the 
 color-spectral type calibration of Bessell (1990). This calibration holds 
 for solar-metallicity stars. For stars with $\leq$ 1/41 of the Sun's 
 metallicity, as is the case for SBS 0335--052, the corresponding spectral 
 types at a given (V--I) color are slightly later. In any case, if stars are 
 responsible for the underlying extended emission, they cannot be older than 
 $\sim$ 10$^8$ yr. However we do not believe the underlying emission to be of 
 stellar origin.
%A power-law drop-off with radius of the underlying light is consistent with 
%the hypothesis of gas ionized by central stellar clusters.  
The irregular, blotchy and filamentary structure of the underlying component 
 argues for a gaseous origin. This hypothesis is further  supported by the 
 colors of the 
 shell structure which are themselves very blue, as seen clearly in 
 the color map in Figure 3. Aperture measurements at various locations along 
 the shell give (V--I)$_0$ between --0.75$\pm$0.20 and --0.18$\pm$0.20. 
The bluest 
 colors are bluer than that of the hottest star ((V--I)$_0$=--0.33 for an OV 
 star according to Bessell (1990)), but can be understood if due to 
 gaseous emission. Another line of evidence is provided by 
 the long-slit observations of Izotov et al. (1996) of the underlying 
 component. Those authors detected strong 
 [OIII]$\lambda$4363 emission indicating high excitation gas as far as 
 $\sim$ 2 kpc from the SSCs. Using the observed 
%fluxes 
equivalent widths of the H$\beta$, 
 [OIII]$\lambda\lambda$ 4959, 5007 and H$\alpha$ lines and continuum fluxes
 from free-free and free-bound 
 processes, and convolving with the V and I bandpasses, they
 were able to reproduce the 
 (V--I)$_0$ color range observed for the underlying component. We note here 
 that these very blue colors are not the results of some observational or 
 reduction procedure artifact. We have obtained HST WFPC2 V and I images of 
 the BCD Mrk 996 in exactly the same observational configuration, 
and reduced them in exactly the same manner (Thuan et al. 1996a). 
 Yet, Mrk 996 shows a red underlying disk with a system of red globular 
 clusters ((V--I)$_0$ $\sim$ 0.9).

\begin{center}
3.3. {\sl A primordial neutral hydrogen intergalactic cloud?}
\end{center}
Here we present indirect evidence which also supports the youth 
hypothesis of SBS 0335--052.
 We use the BCD as a background light source shining through 
 the HI envelope in which it is embedded (Thuan et al. 1996b) to  
probe the physical conditions of the surrounding neutral gas.
 The Ly$\alpha$ line seen in absorption would 
 give the column density of atomic hydrogen, while the OI $\lambda$1302
 absorption line 
 would give the column density of the most abundant heavy element which 
 remains neutral in the HI cloud. This would allow us to set a limit on the 
O/H abundance ratio in the neutral gas.

   The spectrum showing the spectral region 
around the resonance line of oxygen at 1302 \AA\, is shown in Figure 6b.
The OI line is not seen in absorption.  
 We can place an upper limit of $\sim$ 200 m\AA\ (corresponding to 2 times 
 the rms noise) 
 for its equivalent width, 
 which corresponds to a column density of 2.7$\times$10$^{14}$ 
O atom cm$^{-2}$, 
 using the relation given by Kunth et al. (1994) between the equivalent width 
 and the column density for an unsaturated line. Since the HI column density 
 is $\sim$ 10$^{21}$ H atom cm$^{-2}$ from the VLA map 
 (Thuan et al. 1996b), 
the oxygen to hydrogen abundance ratio is $\leq$ 2.7$\times$10$^{-7}$. 
This is more than 3000 
 times smaller the N(O)/N(H) of Orion equal to 8$\times$10$^{-4}$ 
and more than 3 
 times lower than the oxygen abundance measured in the HI envelope of IZw18, 
 where the OI $\lambda$1302 line was clearly detected in absorption 
 ( Kunth et al. 1994). This 
observation provides further support for the youth hypothesis of 
SBS 0335--052. The HI cloud 
 in which SBS 0335--052 is embedded appears to be composed of
primordial pristine 
 gas, uncontaminated by metal-enrichment from previous star formation.

\begin{center}
4. DISCUSSION
\end{center}

 From the (V--I)$_0$ colors of the super-star clusters and using Leitherer 
 \& Heckman (1995)' models, we have concluded that the present burst of star 
 formation did not start until $\leq$ 3--4 Myr ago. From the very blue and 
 filamentary structure of the underlying low-surface-brightness component, we 
 have concluded that the latter has not a  stellar but gaseous origin. Because 
 SBS 0335--052 does not appear to possess an underlying older stellar
 population, the 
 present starburst is probably the first one in the galaxy's history, 
 and the BCD is a good candidate for being a 
 young galaxy. Is this conclusion consistent with other observational 
 constraints?

%\begin{center}
%4.1. {\sl Broad-band colors}
%\end{center}

%We first check that the other observed colors for SBS 0335--052 are also 
% consistent with such a young age. Leitherer \& Heckman (1995)' model with 
% a continuous star formation rate, a normal IMF, Z=Z$_\odot$/10, M$_u$= 100 
% M$_\odot$, M$_l$=
%1 M$_\odot$ and no underlying older stellar population predicts (B--V) $\sim$ 
% --0.06$\div$0.0, (U--B) $\sim$ --1.7$\div$--1.2 and 
%(V--R) $\sim$ 0.1$\div$0.2 for an age 
% $\leq$ 3--4 Myr. This is to be compared to  the observed colors listed in 
% Table 1.
%These colors were obtained through a 12$''$ diameter aperture centered on 
% SSC1, so they refer to the total light of all 6 SSCs. If we adopt a mean 
% extinction of E(B--V) $\sim$ 0.3 mag (Melnick et al. 1992) and the
% extinction 
% curve tabulated by Johnson (1966b), then (B--V)$_0$ $\sim$ 0.0, (U--B)$_0$ 
% $\sim$ --0.63, and (V-R)$_0$ $\sim$ 0.16, in good agreement with the model 
% predictions.

\begin{center}
4.1. {\sl Dust}
\end{center}

We have shown that dust is clearly present in the BCD (see the color map in 
Figure 3). Can it be made in the 
 short time scale implied by the young galaxy scenario? There is some 
 observational evidence that dust formation can occur in a rather short time 
 scale in supernova ejecta. Lucy et al. (1995) have attributed the 
 brightening of SN 1987A in the IR after day $\sim$ 450 to condensation of 
 small silicate grains. The dust condensation efficiency is rather low 
 ( $\leq$ 10$^{-3}$ ), but an efficiency of $\sim$1 is possible if the dust 
 is clumped. Thus, while the physics of dust formation are still not very 
 well known, the SN 1987A observations do not rule out the possibility that 
 enough silicate grains can condense out of supernova ejecta to account for 
 the dust observed in SBS 0335--052.
The BCD has not been detected in any of the four IRAS bands. If we adopt as 
upper limit a flux of 0.5 Jy at 100 microns, then using the formula in 
Sauvage \& Thuan (1994) and adopting their mean value f(100 microns)/
 f(60 microns) $\sim$ 2 for BCDs, 
we derive an upper limit of $\sim$ 1.6 x 10$^5$ M$_\odot$ for the warm dust
 mass. 

\begin{center}
4.2. {\sl The supershell}
\end{center}

We next ask whether the size of the shell is compatible with a  time scale 
 of $\sim$ 4 Myr. The radius of the shell is $\sim$ 380 pc, which 
 qualifies it to be a `supershell' according to the definition of 
 Tenorio-Tagle \& Bodenheimer (1988). To put constraints on the age of the 
 shell structure, we use the models of shell formation of McCray \& Kafatos 
 (1987). These take into account the effects of supernova as the encircled 
 stellar clusters age. We need some estimate of the density of the ambient 
 ISM. The HI mass of the whole cloud (which is much larger than that 
associated with the BCD) is $\sim$ 10$^9$ M$_\odot$, 
 and  the VLA map (Thuan et al. 1996b) shows that most of it is 
 within a projected region 
 31 kpc$\times$7 kpc in size (the whole cloud is larger, extending up to 
 67 kpc$\times$24 kpc, but the outer parts have a very low density 
and do not contribute significantly to the total mass). Assuming a
 cylindrical 
 shape for the HI cloud, an 
 inclination angle of 0 degree and a uniform density gives a mean density 
 of $\sim$ 0.01 H atoms cm$^{-3}$. As SBS 0335--052 is associated with one of 
 the two peaks in HI column density which is $\sim$ 100 times higher than the 
 mean, we obtain a density of $\sim$ 1 H atom cm$^{-3}$ for the ISM around 
 the BCD. This is comparable to the density of $\sim$ 1.5 H atoms cm$^{-3}$ 
 found by Hunter \& Thronson (1995) for the ambient ISM of IZw18. If 
we use the analytical expressions of McCray \& Kafatos (1987) and 
 assume that each supernova explosion produces an energy of 10$^{51}$ ergs,
 then to produce a supershell of radius $\sim$ 380 pc in a time of $\sim$ 4 
Myr, then a total number of stars more massive than 7 M$_\odot$ (all those 
massive enough to undergo a type II supernova) of $\sim$ 14400 is required.
This number is not unreasonable as the total H$\beta$ luminosity ($\sim$ 4
$\times$10$^{40}$ erg s$^{-1}$) gives an equal number of $\sim$ 10000 
O7 stars. The models 
 predict a shell expansion velocity of $\sim$ 56 km s$^{-1}$.

\begin{center}
4.3. {\sl The star formation rate}
\end{center}

%Can the observed amount of metals be produced in the short time available? 
% The HI mass associated with SBS 0335--052 is one order of magnitude 
%smaller than the total mass of the HI cloud, $\sim$ 10$^8$ M$_\odot$ (Thuan 
% et al. 1996b). An O/H ratio of 1/40 the solar value implies that 
%$\sim$ 2$\times$  
% 10$^4$ M$_\odot$ of O must have been processed through
% $\sim$ 5$\times$10$^5$ 
% M$_\odot$ of stars more massive than 10 M$_\odot$, adopting the yield by 
% Weaver \& Woosley (1993). This would then imply the formation of 
%$\sim$ 2$\times$ 
% 10$^6$ M$_\odot$ of stars more massive than 0.1 M$_\odot$, assuming a 
% Salpeter IMF. To produce the amount of metals observed, the average star 
% formation rate (SFR) over a period of $\sim$ 4$\times$10$^6$ yr 
%would have to be 
% $\sim$ 0.5 M$_\odot$ yr$^{-1}$.

 We can obtain an estimate of the SFR from 
 the total H$\alpha$ luminosity. Using Melnick et al. (1992)' measurements, 
 we obtain L(H$\alpha$) $\sim$ 5.2$\times$10$^{40}$ erg s$^{-1}$. 
This is a lower 
 limit as Melnick et al. observed only SSCs 1, 2, 4 and 5,  but not
 the fainter SSCs 
 3 and 6 which do not contribute significantly to the 
total H$\alpha$ flux ( less than 10\%).
 Extrapolating the Salpeter IMF down to stars of mass 0.1 M$_\odot$, 
 the SFR is equal to 7.07$\times$10$^{-42}$ L(H$\alpha$) M$_\odot$ yr$^{-1}$ 
 (Hunter \& Gallagher 1986). Since most of the ionizing photons are produced 
 by 30--60 M$_\odot$ stars with lifetimes $\leq$ 3$\times$10$^6$ yr, 
this is the 
 current instantaneous SFR. We obtain a SFR of $\sim$ 0.4 M$_\odot$ 
 yr$^{-1}$ for the BCD. This derived SFR is one order of magnitude larger 
 than that in IZw18 ( $\sim$ 0.04 M$_\odot$ yr$^{-1}$ ) , and is in the upper 
 range of SFRs derived for BCDs (Fanelli, O'Connell \& Thuan 1988). The total 
stellar mass formed during 4 Myr is thus $\sim$ 1.6$\times$10$^6$ M$_\odot$.
The total oxygen mass produced is $\sim$ 6560 M$_\odot$ (Lequeux et al. 1981).

\begin{center}
4.4. {\sl The Lyman $\alpha$ emission}
\end{center}

To obtain insight into the problem of the non-detection of primeval galaxies 
at high redshifts by Ly$\alpha$ searches, we have obtained with HST a 
GHRS spectrum in the spectral region around Ly$\alpha$ at the redshift of 
SBS 0335--052. The spectrum is shown in Figure 6a.
The most striking feature in this figure is the conspicuous absence of strong 
narrow Ly$\alpha$ emission in SBS 0335--052, just as in IZw18 (Kunth et al. 
 1994), and in 
 contradiction with the predictions of models of young dust-free galaxies 
 undergoing massive star formation (Charlot \& Fall 1993). This problem is 
 not new. Previous Ly$\alpha$ studies of blue compact galaxies ( Deharveng et 
 al. 1985, Hartmann et al. 1988, Terlevich et al. 1993, Giavalisco, Koratkar 
 \& Calzetti 1996) had already shown that Ly$\alpha$ emission is very weak or 
 absent in the UV spectra, despite the extremely strong optical emission 
 lines. The favored explanation for the reduction of Ly$\alpha$ fluxes from 
 recombination values is absorption of multiply scattered Ly$\alpha$ photons 
 by dust in the HII region or in the surrounding HI envelope. This would imply 
 increasing Ly$\alpha$ fluxes in galaxies with decreasing metallicities, 
 since presumably low-metallicity objects contain less dust, and hence suffer 
 less destruction of Ly$\alpha$ photons. The observational evidence is 
 mixed. Terlevich et al. (1993) did find such a correlation, while 
 Giavalisco et al. (1996) found none. The latter authors concluded that the 
 ISM in BCDs is highly inhomogeneous and that the transport of Ly$\alpha$ 
 photons is primarily controlled by the ISM geometry rather than by the dust 
 amount.
The absence of strong narrow Ly$\alpha$ emission in the two most 
 metal-deficient BCDs known, IZw18 (Z$_\odot$/50) and SBS 0335--052 
 (Z$_\odot$/41), supports the lack of correlation between Ly$\alpha$ fluxes 
and metallicities found by 
 Giavalisco et al. (1996).

 In the case of SBS 0335--052, 
 dust absorption may play some role in Ly$\alpha$ photon destruction since 
 dust is clearly present in the BCD 
 (Figure 3). But it may not play the dominant role. Most of the
 light in the 2$''$$\times$2$''$ aperture comes 
 from SSC1 which, as we have argued, does not suffer from extinction. This 
 conclusion is reenforced by the comparison of the observed continuum level 
 near Ly$\alpha$ ($\sim$ 7.5$\times$10$^{-15}$ erg s$^{-1}$ 
\AA$^{-1}$ cm$^{-2}$) 
 with that near H$\beta$ in the same 2$''$$\times$2$''$ aperture 
 (Izotov et al. 1996).
The ratio is $\sim$ 100, close to the value predicted for young dust-free 
 starbursts ( Charlot \& Fall 1993). We favor therefore multiple scattering 
 of Ly$\alpha$ photons in an optically thick gas as the main mechanism of 
 Ly$\alpha$ attenuation. There is observational evidence that the gas 
 surrounding SBS 0335--052 is optically thick for Ly$\alpha$ photons. 
 Figure 6a shows that, contrary to the situation in IZw18 (Kunth et al. 1994) 
, the intensity of 
 the central part of the Ly$\alpha$ absorption caused by the HI envelope does 
 not go to zero. In fact, superposed on the broader absorption,  there is a 
 broad (FWZI=20\AA) low-intensity emission with a total flux of 2.6$\times$
 10$^{-14}$ erg cm$^{-2}$ s$^{-1}$. Broad emission is expected if 
 resonant scattering occurs in a medium with optical depth of $\sim$ 10$^8$ 
 for Ly$\alpha$ photons (Bonilha et al. 1979). This optical depth would 
 correspond to a HI column density of $\sim$ 10$^{21}$ H atoms cm$^{-2}$, 
 which is about the observed value in the VLA map (Thuan et al. 1996b). 
 Multiple scattering 
 redistributes Ly$\alpha$ emission over the whole extent of the HI cloud 
 which is much larger than the aperture used. By using a very small  
 aperture, most of the re-emitted photons are lost. This explains why the 
 Ly$\alpha$-to-H$\beta$ ratio in the BCD ($\sim$ 0.45) is so small 
 compared to the case B recombination value of 33.

 The geometry of the HI cloud may also play a role. In SBS 0335--052, the 
HI cloud is 
 seen nearly edge-on, as suggested by its flattened structure (Thuan et al. 
 1996b).
% and the small velocity gradient ( $\leq$ 33 km s$^{-1}$ kpc$^{-1}$ ) 
% seen in the ionized gas along the SE-NW direction.
 Charlot \& Fall (1993) 
 have shown that, because photons escape more easily in a direction 
 perpendicular to the plane than along it, the ratio of the Ly$\alpha$ 
 intensity to its mean intensity changes from a value of 2.3 for face-on 
 galaxies to 0.0 for edge-on galaxies.

What are the implications of the above observations concerning the 
problem of the non-detection of high-redshift primeval galaxies in Ly$\alpha$ 
searches? Dust is clearly present in a very metal-deficient environment 
such as in SBS 0335--052, so that dust absorption may play a role.
Geometrical effects may also contribute. Galaxies seen face-on are easier 
to detect than those seen edge-on. Also we expect metal-rich galaxies 
 which have 
undergone repeated bursts of star formation to have a more eroded HI envelope 
with more holes through which Ly$\alpha$ photons can escape. They would 
 be more  detectable than young very metal-deficient galaxies undergoing 
 their very first burst of star formation with a less porous HI envelope. 
Lequeux et al. (1995) have discussed another mechanism which allows the 
escape of Ly$\alpha$ photons. They detected attenuated Ly$\alpha$ emission 
 in the 
relatively metal-rich ( Z $\sim$ Z$_\odot$/3 ) BCD Haro 2. Some Ly$\alpha$ 
photons could escape because they are moving at a different velocity from 
that of the surrounding HI gas, the velocity difference 
 ( $\sim$ 200 km s$^{-1}$ ) 
being caused by a galactic wind. In summary, the escape and detection 
 of Ly$\alpha$ photons depends not only on the amount of dust present, 
but also on the geometry and velocity structure of the surrounding HI 
envelope.

\begin{center}
4.5. {\sl Comparison with other nearby young galaxy candidates}
\end{center}

 Several other nearby objects  have been proposed as 
possible young galaxies. This includes IZw18, the only BCD known to be 
more metal-deficient than SBS 0335--052, and which also does not possess a 
low-surface-brightness underlying old stellar population (Kunth et al. 
 1988, Hunter \& Thronson 1995). These two BCDs with no apparent underlying 
older stellar component 
 constitute the exception rather the rule.

 The HI cloud associated with IZw18 is much smaller in intrinsic size ( 6 kpc
$\times$3 kpc, Viallefond, Lequeux \& Comte 1987) 
 than the one in which SBS 0335--052 is embedded. For both BCDs, the 
 starburst region is associated with, but does not coincide exactly with 
the HI column density peak. This offset of $\sim$ 1 kpc is a general 
feature of all BCDs with HI interferometric maps ( Viallefond \& Thuan 1983).
The HI cloud associated with SBS 0335--052 has another HI peak at $\sim$ 1.5
$'$ west ($\sim$ 24 kpc). There is a faint (V $\sim$ 19 mag) compact 
(2.8$''$$\times$2.2$''$) dwarf galaxy at the location of the second HI peak.
Its redshift ( v= 3990$\pm$40 km s$^{-1}$ ) is about the same as that of 
SBS 0335--052. Its spectrum shows H$\alpha$, H$\beta$ and 
 [OIII]$\lambda$5007 emission, but has lower excitation than that of the 
 BCD ( Pustilnik et al. 1996).

 The very large size (64 kpc $\times$ 24 kpc) of the HI cloud of 
 SBS 0335--052 is very unusual for BCDs. The typical size is more like a 
few kpc in each dimension (Viallefond \& Thuan 1983). This large cloud is 
more reminiscent of the intergalactic cloud HI 1225+01 with a size of 
 $\sim$ 100 kpc (adopting a distance of 10 Mpc, Salzer et al. 1991). 
Both clouds have HI masses of $\sim$ 10$^9$ M$_\odot$. There are also 2 HI 
peaks in HI 1225+01 separated by $\sim$46 kpc, one of which is associated 
with a faint ( M$_B$ $\sim$ --14.0 ) and metal-deficient  
( Z $\sim$ Z$_\odot$/18 ) dwarf irregular galaxy. As for SBS 0335--052, the 
 galaxy is very blue ( B--V= 0.10, U--B= --0.57 ) and most ( $\geq$ 72 
percent ) of the B light comes from stars with ages $\leq$ 40 Myr.
Thus, although the burst in SBS 0335--052 is younger than in HI 1225+01 , 
their HI envelopes have remarkably similar properties. It is not yet clear 
why star formation has  
been stunted in these HI clouds for so long. In the case of SBS 0335--052, a 
 mean density of  $\sim$ 0.01 H atom cm$^{-3}$ leads to a collapse time of
 only $\sim$ 5$\times$10$^8$ yr, much shorter than the Hubble time.

\begin{center}
5. SUMMARY
\end{center}

  We have analyzed WFPC2 V and I images and GHRS UV spectra obtained with 
the refurbished HST of the extremely metal-deficient ( Z = Z$_\odot$/41 ) 
blue compact dwarf (BCD) galaxy SBS 0335--052. We have obtained the following 
results:
 
1) The underlying extended low-surface-brightness component of the BCD is 
extremely blue (--0.34 $\leq$ (V--I)$_0$ $\leq$ 0.16 ). Its color profile 
 is nearly flat. If stars are responsible for this low-surface-brightness 
emission, they cannot be older than $\sim$ 10$^8$ yr. However, the 
 irregular and filamentary structure of the underlying component 
 supports the case for gaseous emission and no underlying older stellar 
population. The very blue colors are  
consistent with gaseous emission.
Thus SBS 0335--052 is very likely a  nearby young galaxy in the sense that it
 is undergoing its very first burst of star formation. 
This conclusion supports the prediction of Izotov et al. (1990)
 about the youth of
SBS 0335--052 on the basis of its very low metallicity.
%The youth hypothesis 
%of the BCD has been discussed previously by Izotov et al. (1990) on the basis 
%of its very low metallicity.

2) The OI $\lambda$1302 line is not detected in absorption in the GHRS 
spectrum, setting an upper limit for N(O)/N(H) in the HI envelope of 
$\sim$ 3$\times$10$^{-7}$, more than 3000 times smaller than the value 
in Orion and more than 3 times lower than the oxygen abundance measured 
in the HI envelope of IZw18, and consistent with the youth hypothesis 
of SBS 0335--052.      

3) All star formation occurs in six very blue super-star clusters (SSC) 
 with ages $\leq$ 3--4 Myr. SSC formation appears to be related to  
starburst events in some, but not all, BCDs. The triggering of star 
formation in SBS 0335--052 is not due to tidal interactions, but is 
probably caused by stochastic collisions of gas clouds within the large 
HI envelope.

4) Dust is clearly present and is spatially mixed with the
 SSCs. While the brightest SSC does not suffer from extinction, there is 
a systematic increase in reddening away from it. Such dust can probably 
condense out in supernova remnants in a time scale $\leq$ 4 Myr.

5) There is a supershell of radius $\sim$ 380 pc delineating a large 
supernova cavity. Its observed properties can be accounted for by 
models of shell formation with a time scale of $\sim$ 4 Myr.

6) The star formation rate (SFR) is $\sim$ 0.4 M$_\odot$ yr$^{-1}$, in the 
upper range of SFRs observed in BCDs.

7) Strong narrow Ly$\alpha$ emission is not detected. There is instead 
low-intensity broad (FWZI= 20\AA) Ly$\alpha$ emission superposed on 
broader Ly$\alpha$ absortion from the HI envelope. The low-intensity 
broad emission is caused by resonant scattering in a dense ambient 
medium, with optical depth of $\sim$ 10$^8$ for Ly$\alpha$ photons, 
corresponding to the observed HI column density of $\sim$ 10$^{21}$ 
H atom cm$^{-2}$.

While dust may play some role in the destruction of 
 Ly$\alpha$ photons, multiple scattering which removes Ly$\alpha$ 
photons from the small HST aperture and redistributes them over the whole 
HI cloud, and geometrical effects ( the HI cloud is seen edge-on) appear to
 be the main mechanisms for removal of Ly$\alpha$ photons in SBS 0335--052.
 Ly$\alpha$ detection 
of high-redshift primeval galaxies depends thus not only on the amount of 
dust, but also on the geometry and velocity structure of the surrounding 
HI envelope.

\vspace{1.0cm}

T.X.T. acknowledges the partial financial support of STScI grant 
 GO 5408.01-93A. Y.I.I. is grateful for the hospitality 
 of the Astronomy 
 Department of the University of Virginia. This international collaboration 
 was made possible by NATO collaborative research grant 921285 and INTAS
international grant 94-2285.

\begin{center} 
REFERENCES 
\end{center}

%\section*{References}
\begin{description}
  \itemsep-.8ex
\item{Baron, E., \& White, S.D., 1987, ApJ, 322, 585}
\item{Barth, A.J., Ho, L.C., Fillipenko, A.V., \& Sargent, W.L.W. 1995, AJ,
110, 1009}
\item{Bessell, M.S. 1990, PASP, 102, 1181}
\item{Bonilha, J.R.M., Ferch, R., Salpeter, E.E., Slater,G. \& Noerdlinger, 
P.D. 1979, ApJ, 233, 649}
\item{Burstein, D., \& Heiles, C. 1982, AJ, 87, 1165}
%\item{Cervi\~{n}o, M., \& Mas-Hesse, J.M. 1994, A\&A, 284, 749}
\item{Charlot, S., \& Fall, S.M. 1993, ApJ, 415, 580}
\item{Conti, P.S., \& Vacca, W.D. 1994, ApJ, 423, L97}
%\item{Danziger, I.J. 1983, in Supernova Remnants and their X-Ray Emission,
%     ed. J.Danziger \& P.Gorenstein, 193}
\item{Deharveng, J.-M., Joubert, M. \& Kunth, D. 1985, in Star-forming Dwarf 
 Galaxies and related objects, ed. D. Kunth, T.X. Thuan \& J.T.T. Van, 
 (Gif-sur-Yvette:Editions Frontieres), 431}
%\item{Denisyuk, E.K. \& Lipovetsky V.A., 1984, Astrophysics, 20, 290}
\item{Fanelli, M.N., O'Connell, R.W. \& Thuan, T.X. 1988, ApJ, 334, 665}
%\item{Feibelman, W.A., Aller, L.H., \& Hyung, S. 1992, PASP, 104, 339} 
%\item{Ferland, G.J. 1993, University of Kentucky Department of Physics and
%     Astronomy Internal Report}
\item{Fontana, A., Cristiani, S., D'Odorico, S., Giallongo, E., \& Savaglio, S.
1996, MNRAS, in press}
%\item{Freeman, K.C. 1970, ApJ, 160, 811}
%\item{Garnett, D.R., Skillman, E.D., Dufour, R.J., Peimbert, M., 
%    Torres-Peimbert, S., Terlevich, R., Terlevich, E., \& Shields, G.A. 1995,
%    ApJ, 443, 64}
\item{Giavalisco, M., Koratkar, A., \& Calzetti, D. 1996, STSI preprint 
 No.1021}
%\item{Harris, W.E. 1991, ARA\&A, 29, 543}
%\item{Harris, H.C., Hunter, D.A., Baum, W.A., \& Jones, J.H. 1993, AJ, 105,
%    1196}
\item{Hartmann, L.W., Huchra, J.P., Geller, M.J., O'Brien, P. \& Wilson, R. 
 1988, ApJ, 236, 101}
\item{Ho, L.C. 1996, Rev. Mex. Astr. Astrofis., in press}
\item{Holtzman, J. et al. 1992, AJ, 103, 691}
\item{Holtzman, J.A., Burrows, C.J., Casertano, S., Hester, J.J., Trauger, 
 J.T.,
    Watson, A.M., \& Worthey, G. 1995, PASP, 107, 1065}
\item{HST WFPC2 Handbook 1995}
\item{Hunter, D.A., \& Gallagher,J.S. 1986, PASP, 98, 5}
\item{Hunter, D.A., O'Connell, R.W., \& Gallagher, J.S. 1994, AJ, 108, 84}
\item{Hunter, D.A., \& Thronson, H.A., Jr. 1995, ApJ, 452, 238}
%\item{Hyung, S., Aller, L.H., \& Feibelman, W.A. 1994, ApJS, 93, 465}
\item{Izotov, Yu.I., Lipovetsky, V.A., Guseva, N.G., \& Kniazev, A.Yu. 1992, 
 in
    The Feedback of Chemical Evolution on the Stellar Content of Galaxies, 
 ed. 
D. Alloin \& G.Stasinska (Paris Observatory Publ.), 138}
\item{Izotov, Yu.I.,Guseva, N.G., Lipovetsky, V.A., Kniazev, A.Yu.,
    Neizvestny, S.I., \& Stepanian, J.A., 1993, Astron.Astroph.Trans. 3, 197}
\item{Izotov, Yu.I., Lipovetsky, V.A., Chaffee, F., Foltz, C., Guseva, N.G.,
\& Kniazev, A.Yu. 1996, in preparation}
\item{Izotov, Yu.I., Lipovetsky, V.A., Guseva, N.G., Kniazev, A.Yu., \&
Stepanian, J.A. 1990, Nature, 343, 238}
\item{Izotov, Yu.I., Thuan, T.X., \& Lipovetsky, V.A. 1994, ApJ, 435, 647}
\item{Izotov, Yu.I., Thuan, T.X., \& Lipovetsky, V.A. 1996, ApJ, submitted}
%\item{Johnson, H.L. 1966a, ARA\&A, 4, 193}
%\item{Johnson, H.L. 1966b, in Nebulae and Interstellar Matter, ed. B.M. 
%Middlehurst \& L.H.Aller (Chicago: University of Chicago Press), 167}
%\item{Kingdon, J., \& Ferland, G.J. 1995, ApJ, 442, 714}
\item{Kunth, D., \& Sargent, W.L.W. 1986, ApJ, 300, 496}
\item{Kunth,D., Maurogordato,S. \& Vigroux,L. 1988, A\&A, 204,10}
\item{Kunth, D., Lequeux, J., Sargent, W.L.W., \& Viallefond, F. 1994, A\&A,
282, 709}
\item{Landolt, A.U. 1992, AJ, 104, 340}
\item{Leitherer, C., \& Heckman, T.M. 1995, ApJS, 96, 9}
\item{Lequeux, J., Maucherat-Joubert, M., Deharveng, J.-M., \& Kunth, D. 1981,
A\&A, 103, 305}
\item{Lequeux, J., Kunth, D., Mas-Hesse, J.M., \& Sargent, W.L.W. 1995,
A\&A, 301, 18}
%\item{Lipovetsky, V.A., Thuan, T.X., Richter, G.M., Pustilnik, S.A., \&
%     Izotov, Yu.I. 1996, in preparation}
\item{Loose, H.-H., \& Thuan, T.X. 1985, in Star-Forming Dwarf galaxies and 
related objects, ed. D. Kunth, T.X.Thuan \& J.T.T.Van (Gif-sur-Yvette: 
 Editions 
Frontieres), 73}
%\item{Loose, H.-H., \& Thuan, T.X. 1986, ApJ, 309, 59}
\item{Lucy, L.B., Danziger, I.J., Gouiffes, C. \& Bouchet, P. 1991, in 
Supernovae, ed. S.E.Woosley (New York: Springer-Verlag), 82}
%\item{Lutz, J.H., Kaler J.B., Shaw, R.A., Schwarz, H.E., \& Aspin, C. 1989,
%PASP, 101, 966}
%\item{Maeder, A., \& Meynet, G. 1994, A\&A, 287, 803}
\item{Markarian, B.E., Lipovetsky, V.A., \& Stepanian, J.A. 1983, Astrofizika, 
     19, 29}
\item{McCray, R. \& Kafatos, M. 1987, ApJ, 317, 190}
\item{Meier, D.I. 1976, ApJ, 207, 343}
\item{Melnick, J., Heydari-Malayeri, M., \& Leisy, P. 1992, A\&A, 253, 16}
\item{Meurer, G.R., Heckman, T.M., Leitherer, C., Kinney, A., Robert, C., \&
Garnett, D.R. 1995, AJ, 110, 2665}
\item{O'Connell, R.W., Gallagher, J.S., \& Hunter, D.A. 1994, ApJ, 433, 65}
\item{Papaderos, P., Loose, H.-H., Thuan, T.X. \& Fricke,K.J. 1996,
A\&AS, in press}
\item{Partridge, R.B., \& Peebles, P.J.E. 1967, ApJ, 147, 868}
\item{Pettini, M., Lipman, K., \& Hunstead, R.W. 1995, ApJ, 451, 100}
%\item{Pettini, M., Smith, L.J., Hunstead, R.W., \& King, D.L. 1994, ApJ, 426, 
%79}
\item{Pritchett, C.J. 1994, PASP, 106, 1052}
%\item{Prochaska, J.X., \& Wolfe, A.M. 1996, astro-ph/9604042}
\item{Pustilnik, S.A., Ugryumov, A.V., Lipovetsky, V.A., Thuan, T.X. \&
     Guseva, N.G. 1995, ApJ, 443, 499}
\item{Pustilnik, S.A., Lipovetsky, V.A., Izotov, Yu.I., Brinks, E.,
Thuan, T.X., Kniazev, A. Yu., Neizvestny, S.I., \& Ugryumov, A.V. 1996, 
Soviet AJ, submitted} 
%\item{Rocca-Volmerange, B., Pr\'{e}vot, L., Ferlet, R., Lequeux, J. \&
%     Pr\'{e}vot-Burnichon, M.L. 1981, A\&A, 99, L5}
\item{Salzer, J.J. 1989, ApJ, 347, 152}
\item{Salzer, J.J., Alighieri, S.D.S., Matteuci, F., Giovanelli, R. \& 
Haynes, M.P. 1991, AJ, 101, 1258}
%\item{Schombert, J.M., Pildes, R.A., Eder, J.A., \& Oemler, A. Jr. 1995,
%     preprint}
\item{Sauvage,M. \& Thuan,T.X. 1994, ApJ, 429,153}
\item{Searle, L., \& Sargent, W.L.W. 1972, ApJ, 173, 25}
\item{Searle, L., Sargent, W.L.W., \& Bagnuolo, W.G. 1973, ApJ, 179, 427}
%\item{Smith, L.J., Pettini, M., King, D.L., \& Hunstead, R.W. 1996,
%astro-ph/9601153}
%\item{Steidel, C.C., Giavalisco, M., Pettini, M., Dickinson, M., \&
%Adelberger, K.L. 1996, ApJ, in press}
%\item{Tenorio-Tagle, G. 1994, in Violent Star Formation from 30 Doradus to
%QSOs, ed. G. Tenorio-Tagle (Cambridge: Cambridge Univ. Press), 50}
\item{Tenorio-Tagle, G. \& Bodenheimer, P. 1988, ARAA, 26,145}
\item{Terlevich, E., Diaz, A.I., Terlevich,R. \& Garcia Vargas, M.L. 1993, 
MNRAS, 260,3 }
\item{Terlevich, R., Melnick, J., Masegosa, J., Moles, M., \& Copetti, M.V.F.
1991, A\&AS, 91, 285}
\item{Thuan, T.X. 1983, ApJ, 268, 667}
\item{Thuan, T.X. 1991, in Massive Stars in Starbursts, ed. C.Leitherer,
 N.R. Walborn, T.M. Heckman \& C.A. Norman,
(Cambridge: Cambridge University Press), 183}
\item{Thuan, T.X., \& Martin, G.E. 1981, ApJ, 247, 823}
%\item{Thuan, T.X. \& Sauvage, M. 1992, A\&AS, 92, 749}
%\item{Thuan, T.X., Izotov, Yu.I., Lipovetsky, V.A. \& Pustilnik, S.A. 1994,
%in Dwarf Galaxies, ed. G. Meylan \& P. Prugniel (Garching: European 
%Southern Observatory), 421}
\item{Thuan, T.X., Izotov, Yu.I., \& Lipovetsky, V.A. 1995, ApJ, 445, 108}
\item{Thuan, T.X., Izotov, Yu.I., \& Lipovetsky, V.A. 1996a, ApJ, 463, 120} 
\item{Thuan, T.X., Brinks, E., Pustilnik, S.A., Lipovetsky, V.A., \& 
Izotov, Yu.I. 1996b, in preparation}
\item{Thuan, T.X., Pustilnik, S.A., Martin, J.-M., \& Lipovetsky, V.A. 
     1996c, in preparation}
%\item{Vacca, W.D., \& Conti, P.S. 1992, ApJ, 401, 543}
\item{Viallefond, F. \& Thuan, T.X. 1983, ApJ, 269, 444}
\item{Viallefond, F., Lequeux, J. \& Comte, G. 1987, in Starbursts and 
Galaxy Evolution, ed. T.X. Thuan, T. Montmerle \& J.T.T. Van 
(Gif-sur-Yvette: Editions Frontieres), 139} 
\item{ Weaver, T.A., \& Woosley, S.E. 1993, Phys. Rep., 227, 65}
\item{Whitford, A.E. 1958, AJ, 63, 201}
\item{Whitmore, B.C., \& Schweizer, F. 1995, AJ, 109, 960}
\item{Whitmore, B.C., Schweizer, F., Leitherer, C., Borne, K., \& Robert, C.
1993, AJ, 106, 1354}
%\item{Womble, D.S., Sargent, W.L.W., \& Lyons, R.S. 1995, astro-ph/9511035}
\item{Yee, H.K.C., Ellington, E., Bechtold, J., Carlberg, R.G., \& Cuillandre,
J.-C. 1996, AJ, 111, 1783}

\end{description} 
\newpage
\begin{center}
FIGURE CAPTIONS
\end{center}

   Fig.1. (a) A 1800 s $V$ WFPC2 image of the blue compact dwarf galaxy 
SBS 0335--052 taken with the refurbished HST, with the contrast adjusted 
to show the high surface brightness super-star clusters (SSCs). At a distance 
of 54.3 Mpc, 1$''$ corresponds to a linear size of 263 pc. The orientation 
is the same as in Figure 1b.
(b) A $V$ contour map showing the labeling of the six SSCs listed in Table 2. 
A supershell delineating a large supernova cavity of radius $\sim$ 380 pc 
 is clearly 
seen at the NE end of the galaxy.

  Fig.2. The (V--I)$_0$ color corrected for foreground extinction of the six 
super-star clusters (SSCs) listed in Table 2 as a function of their separation 
from SSC1. There is a systematic increase in reddening with increasing 
distance from SSC1, along the SE-NW direction.

  Fig.3. (V--I) grey scale color map. Blue is dark and red is light. The 
 supershell is very blue while the red dust patches are seen to increase in 
 importance away from super-star cluster 1 in the SE-NW direction. Two 
 background red galaxies can be seen to the south of SBS 0335--052.

 Fig.4. The same V image as in Figure 1a but with the contrast adjusted to 
 show the low-surface-brightness underlying component of SBS 0335--052. The 
latter has a blotchy and irregular appearance, suggesting that the light is 
not emitted by stars but by ionized gas. The supershell delineating the large 
 supernova cavity is clearly seen.

 Fig.5. (a) and (b) $V$ and $I$ surface brightness profiles as a function of 
 log r where r is the equivalent radius. A power law 
 is a straight line in this coordinate system. The error bars take into 
account photon statistics. The large error bars in the radius range --0.45 
 $\leq$ log r $\leq$ --0.05 are caused by light contamination from the 
shell structure.
 (c) (V--I) color profile. It is approximately flat and shows an extraordinary 
blue color for the underlying low-surface-brightness component.

 Fig.6.  GHRS spectra obtained through a 2$''$$\times$2$''$ rectangular 
aperture 
with HST and centered on super-star cluster 1. The aperture includes also 
super-star clusters 2 and 3. The spectral resolution is 0.12 \AA. (a) shows 
the spectral region around Ly$\alpha$. The strong geocoronal Ly$\alpha$ line 
 can be seen at 1216 \AA. The redshifted location of the Ly$\alpha$ line in 
SBS 0335--052 (v = 4043 km s$^{-1}$) is marked. There is low-intensity 
broad (from 1220\AA\ to 1240\AA) Ly$\alpha$ emission superposed on even
 broader absorption. The continuum level is 7.5$\times$10$^{-15}$
 erg cm$^{-2}$ s$^{-1}$ 
\AA$^{-1}$. (b) shows the spectral region around the OI $\lambda$1302 line.
Its redshifted location is marked. The line is not detected and an upper 
limit of $\sim$ 200 m\AA\ can be set for its equivalent width.       

\newpage
 \begin{center}
Table 1. Observational characteristics of SBS 0335--052
\end{center}

%\vspace{0.2cm} 

\begin{center}
 \begin{tabular}{lr  } \hline \hline
 Parameter ~~~~~~~~~~~~~~~~~~~~~~~~~~~~~~~~& Value ~~~~               \\ \hline \\
$\alpha$(1950)\ ............................     & 03$^h$35$^m$15.2$^s$        \\
$\delta$(1950)\ .............................    & --05$^\circ$12$'$26$''$        \\
l$^{II}$, b$^{II}$\ .............................. & 193$^\circ$, --45$^\circ$    \\
d$_{25}$$^a$\ ................................... & 12.6$''$                    \\
D (Mpc)$^b$\ ..........................            & 54.3                        \\
d (kpc)\ ..............................           &  3.29                       \\
V$^a$\ .....................................     & 16.65 $\pm$0.01               \\
%U-B$^c$\ ..................................       & --0.42 $\pm$0.09              \\
%B-V$^c$\ ..................................       & 0.32 $\pm$0.03                \\
%V-R$^c$\ ..................................       & 0.65 $\pm$0.04                \\
V-I$^a$\,\ ..................................    & --0.23$\pm$0.01                        \\
M$_B$\,\ ....................................     & --16.70                      \\
%$\mu_V$ (mag/$\Box ^")^d$\ ...................    &  19.90                      \\
%$\mu_I$ (mag/$\Box ^")^d$\ ....................   & 19.05                       \\
%$\alpha^{-1}$ (kpc)$^d$\ .........................& 0.42                        \\
v$_{HI}$ (km s$^{-1})^c$\,\ ...................    & 4043 $\pm$10                \\
$\Delta v_{20}$ (km s$^{-1})^c$\ ..................&  105 $\pm$14                \\
$\Delta v_{50}$ (km s$^{-1})^c$\ ..................&   83 $\pm$ 9                \\
F$_{HI}$ (10$^6$ M$_\odot$ Mpc$^{-2}$)$^c$\ .......& 0.34                        \\
M$_{HI}$ (10$^9$ M$_\odot)^c$\ .................. & 0.99                        \\
L$_{B} $ (10$^9$ L$_\odot)^d$\ ......................  & 1.1                         \\
M$_{HI}$/L$_B$ (M$_\odot$/L$_\odot$)\ ..............& 0.90                   \\ \hline

 \end{tabular}
 \end{center}

$^a$ from this paper. d$_{25}$ is derived from Figure 2. V and I measured
in a 12$''$ diameter aperture centered on the brightest knot.

\vspace{0.2cm}

$^b$ redshift distance corresponding to H$_0$ = 75 km s$^{-1}$Mpc$^{-1}$ 
and a systematic velocity of 4076 km s$^{-1}$, corrected to the Local Group 
velocity centroid following Thuan \& Martin (1981).

%\vspace{0.2cm}

%$^c$ photoelectric photometry by Izotov et al. (1993) in a 12$''$ diameter
%aperture centered on the brightest knot.

\vspace{0.2cm}

$^c$ HI measurements obtained with the Nan\c{c}ay radio telescope by Thuan 
et al. (1996); v$_{HI}$ is the heliocentric velocity.

\vspace{0.2cm}

$^d$ Adopting M$_B$(sun) = 5.48 mag.

\newpage
 \begin{center}
 Table 2. Magnitudes and colors of stellar clusters
\end{center}

%\vspace{0.2cm} 

\begin{center}
 \begin{tabular}{lccc} \hline \hline

No.&$V_0^a$&$I_0^a$& ($V-I$)$_0^a$ \\ \hline \\

 \  1\ .......... &18.93$\pm$ 0.01&19.54$\pm$ 0.01&--0.61$\pm$ 0.02 \\
 \  2\ .......... &19.24$\pm$ 0.02&19.63$\pm$ 0.02&--0.39$\pm$ 0.02 \\
 \  3\ .......... &20.55$\pm$ 0.08&20.81$\pm$ 0.10&--0.26$\pm$ 0.12 \\
 \  4\ .......... &19.71$\pm$ 0.02&19.86$\pm$ 0.02&--0.15$\pm$ 0.03 \\
 \  5\ .......... &19.50$\pm$ 0.03&19.36$\pm$ 0.03& ~\,0.14$\pm$ 0.04 \\
 \  6\ .......... &21.96$\pm$ 0.11&21.65$\pm$ 0.09& ~\,0.31$\pm$ 0.14 \\ \\
% \  7\ .......... &443.35&434.52&21.373$\pm$ 0.041&21.710$\pm$ 0.036&--0.337$\pm$ 0.055 \\ \\
 \hline
 \end{tabular}
 \end{center}

%$^a$the coordinate system is shown in Figure 2.

%\vspace{0.3cm}

\hspace{2.cm} $^a$corrected for foreground extinction (A$_V$=0.10 mag, A$_I$=

\hspace{2.2cm} 0.06 mag, Burstein \& Heiles 1982; Whitford 1958).

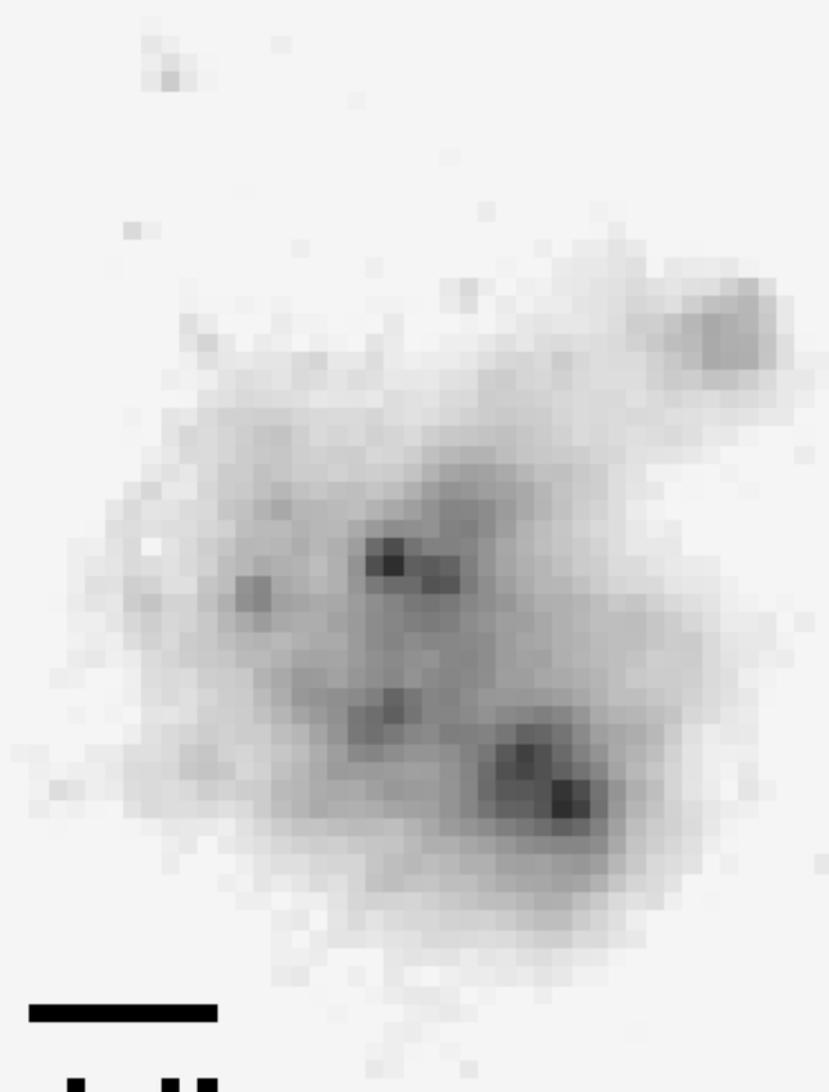
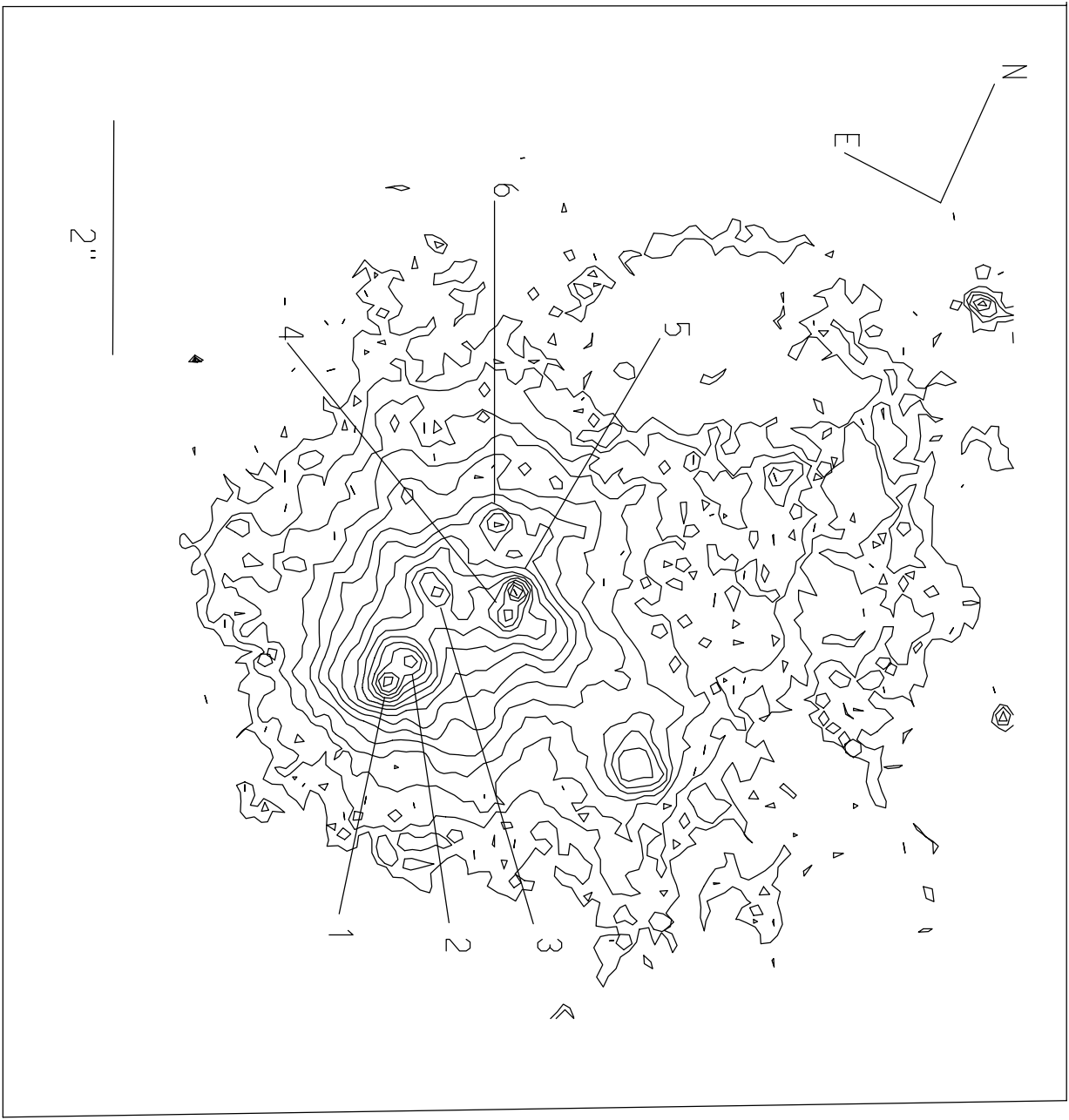
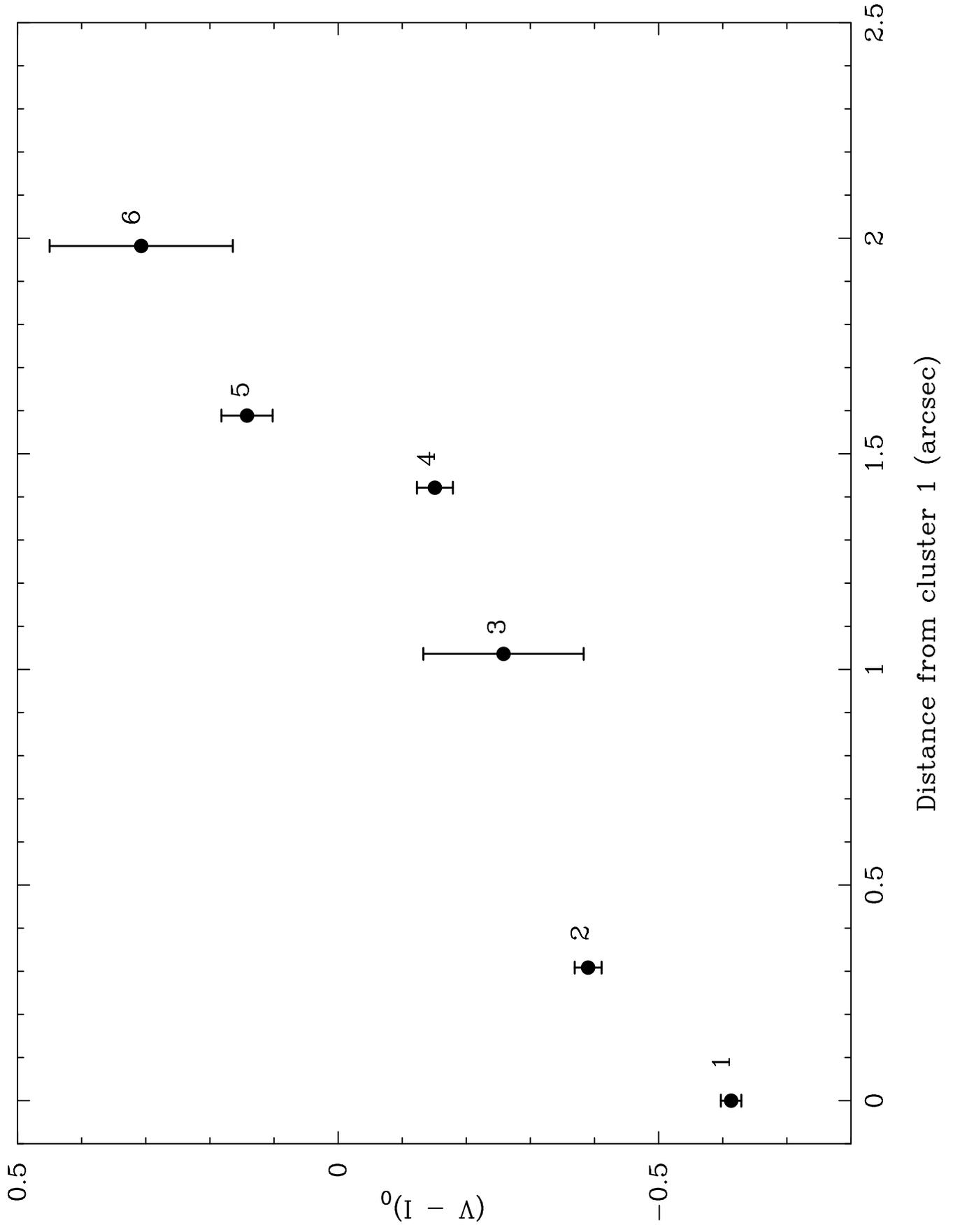
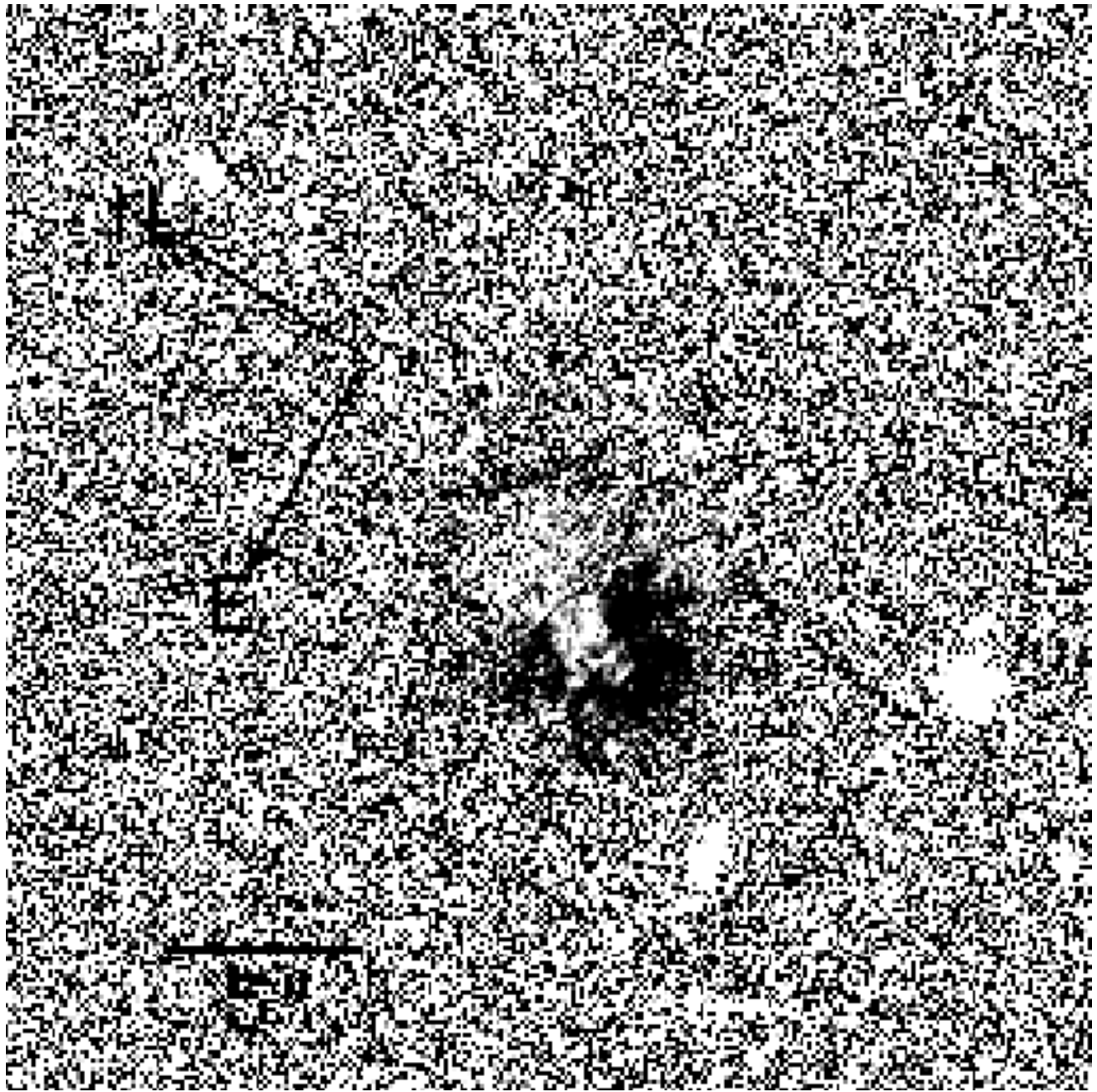
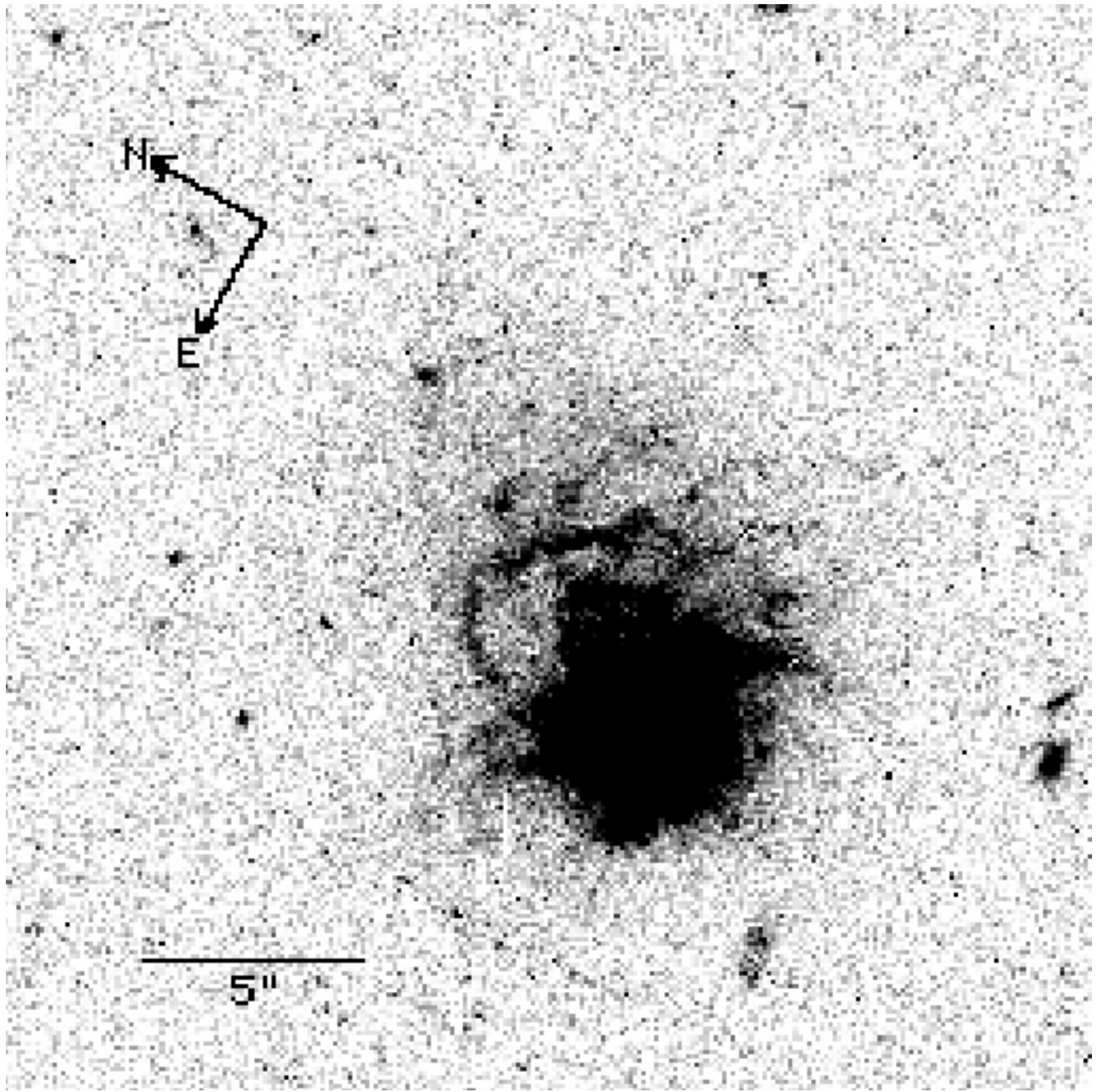
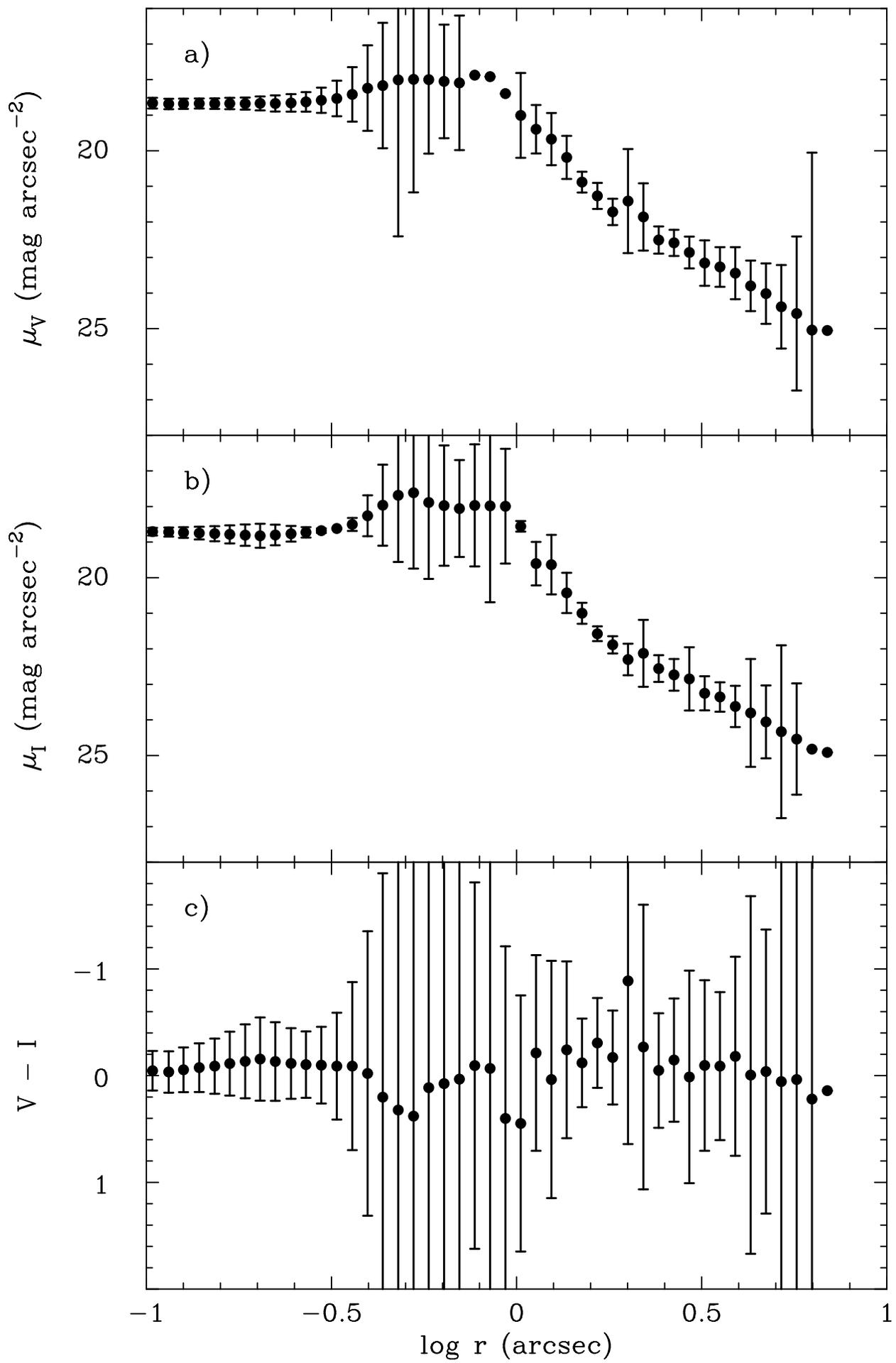
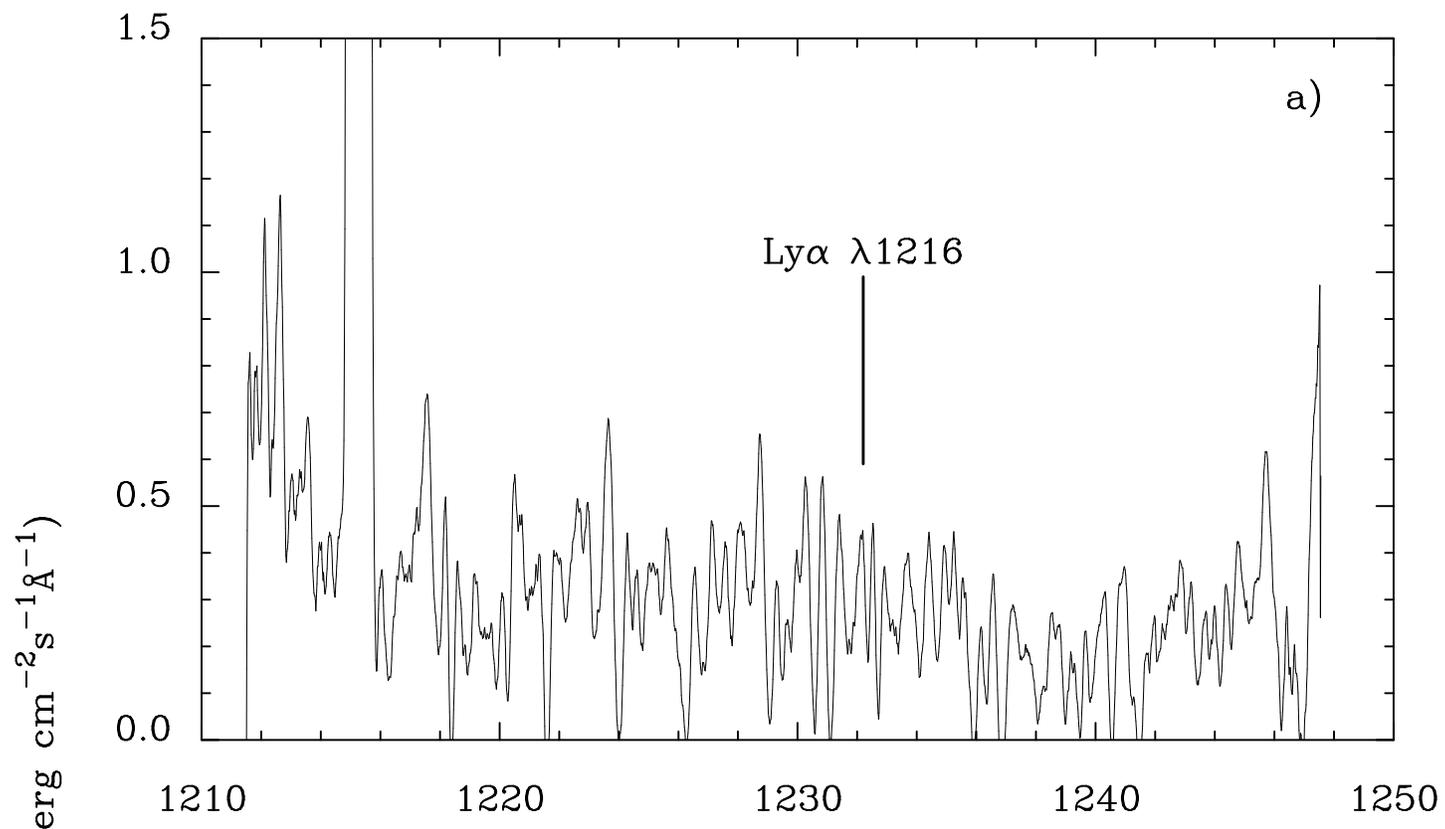
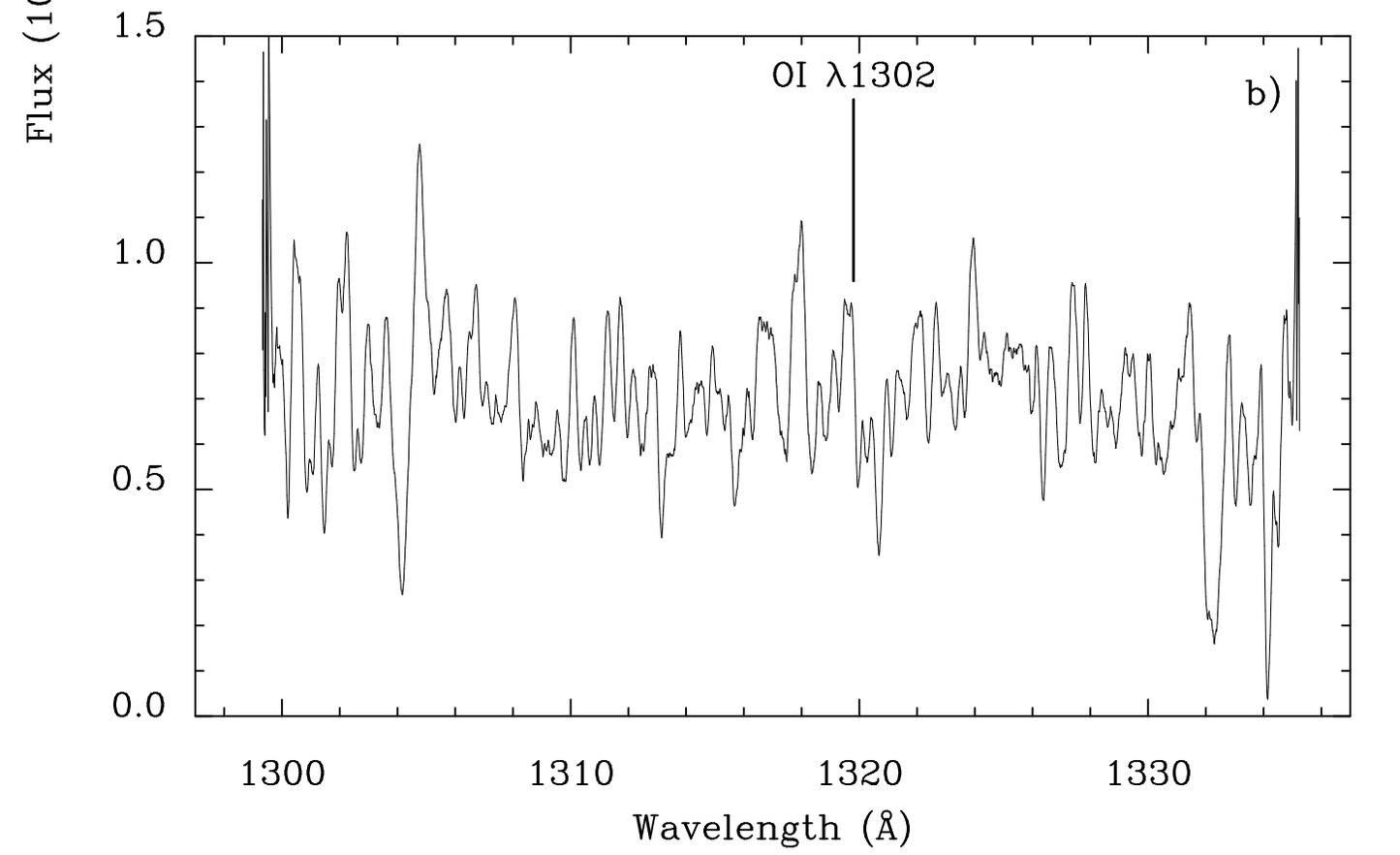
\end{document}